\documentclass[a4paper]{PoS}
\pdfoutput=1
\newcommand{\lapprox}{\raisebox{-0.5ex}{$\
\stackrel{\textstyle<}{\textstyle\sim}\ $}}

\title{Graphene as a Lattice Field Theory}

\ShortTitle{Lattice Graphene}

\author{\speaker{Simon Hands}\\
        Department of Physics, College of Science,
        Swansea University,\\ Singleton Park, Swansea SA2 8PP, U.K.\\
        E-mail: \email{s.hands@swan.ac.uk}}

\author{Wes Armour\\
         Oxford e-Research Centre, University of Oxford, \\7 Keble
         Road, Oxford OX1 3QG, U.K.\\
         E-mail: \email{wes.armour@oerc.ox.ac.uk}}

\author{Costas Strouthos\\
School of Sciences, Department of Computer Science, European University
Cyprus,\\
1516 Cyprus.\\
         E-mail: \email{Strouthos@ucy.ac.cy}}

\abstract{We introduce effective field theories for the electronic properties of
graphene in terms of relativistic fermions propagating in 2+1 dimensions, and
outline how strong inter-electron interactions may be modelled by numerical
simulation of a lattice field theory. For strong enough coupling an 
insulating state can form via condensation of particle-hole pairs, and it is
demonstrated that this is a theoretical possibility for monolayer graphene. 
For bilayer graphene the effect of an interlayer bias voltage can be modelled
by the introduction of a chemical potential 
(akin to isopsin chemical potential in QCD) with no accompanying sign problem;
simulations reveal the presence of strong interactions among the residual
degrees of freedom at the resulting Fermi
surface, which is disrupted by an excitonic condensate. We also present
preliminary results for the quasiparticle dispersion, which permit direct
estimates of both the Fermi momentum $k_F$ and the induced gap $\Delta$.}

\FullConference{9th International Workshop on Critical Point and Onset of Deconfinement - CPOD2014,\\
		17-21 November 2014\\
		ZiF (Center of Interdisciplinary Research), University of Bielefeld, Germany}

\begin{document}

In this talk I will discuss an effective field theory for electron
excitations in graphene, and show how it may be approached using the same lattice
field theory methods more usually applied to the strong interaction between quarks and
gluons. I will argue that for sufficiently strong inter-electron coupling, of the same order
as that found in suspended graphene samples, there is a phase transition to a
insulating phase, described by a quantum critical point (QCP). I will then
specialise to the case of bilayer graphene with a biassing voltage
applied between the layers, so that there is a non-zero density of electrons on
one layer and holes on the other. This is formally very similar to the case of
non-zero chemical potential for isopsin in QCD, and will permit numerical 
simulations probing degenerate matter (albeit in 2+1$d$) in the presence
of strong interactions.

\section{Relativity in Graphene}
Let's begin with a simple tight-binding model, with a Hamiltonian describing
electrons in $\pi$-orbitals hopping between $A$ and $B$ sublattices on the bipartite honeycomb lattice
appropriate
for a graphene monolayer (an excellent introduction can be found in the review
by Castro Neto {\it el al\/}~\cite{CastroNeto:2009zz}):
\begin{figure}[thb]
\vspace{-1cm}
\begin{centering}
\includegraphics[width=.4\textwidth]{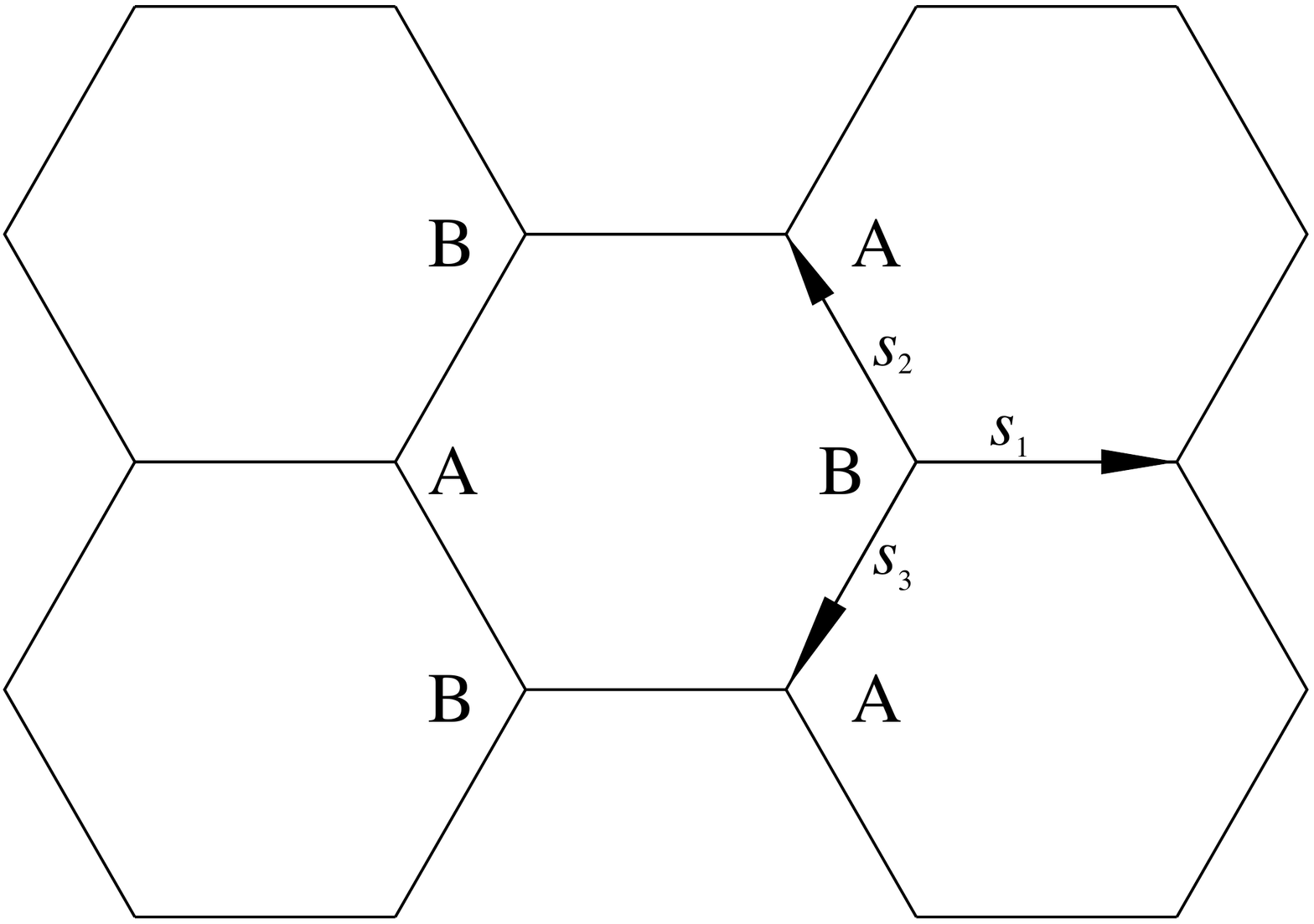}
\vspace{-1cm}
\caption{The honeycomb lattice of monolayer graphene.}
\label{fig:honeycomb}
\end{centering}
\end{figure}
\begin{equation}
H=-t\sum_{\vec r\in B}\sum_{i=1}^3
b^\dagger({\vec r})a({\vec r}+{\vec s}_i)+
a^\dagger({\vec r}+{\vec s}_i)b({\vec r})
\label{eq:tight-binding}
\end{equation}
In momentum space the Hamiltonian is rewritten 
\begin{equation}
H=\sum_{\vec k}\left(\Phi(\vec k)a^\dagger(\vec k)b(\vec k)+
\Phi^*(\vec k)b^\dagger(\vec k)a(\vec k)\right)\;\mbox{\rm with}\;\,
\Phi(\vec
k)=-t\left[e^{ik_xl}+2\cos\Bigl({{\sqrt{3}k_yl}\over2}\Bigr)e^{-i{{k_xl}\over2}}\right]
\end{equation}
where $l$ is the length of a CC bond.
Notice how the fact that a link emerging from a site is not matched by another
leaving at 180$^\circ$ results in a complex-valued kernel $\Phi(\vec k)$. If we
define states $\vert
\vec k_\pm\rangle=(\surd{2})^{-1}[a^\dagger(\vec k)\pm b^\dagger(\vec
k)]\vert0\rangle$, then
\begin{equation}
\langle\vec k_\pm\vert H\vert\vec k_\pm\rangle
=\pm(\Phi(\vec k)+\Phi^*(\vec k))\equiv\pm E(\vec k),
\end{equation}
so that the energy spectrum is symmetric about $E=0$.

In fact, $E=0$ at precisely two independent points within the first Brillouin Zone,
the so-called {\it Dirac points\/} $\vec k=\vec
K_\pm=(0,\pm4\pi/3\surd3 l)$. Linearise about these points;
\begin{equation}
\Phi(\vec K_\pm+\vec p)=\pm v_F[p_y\pm ip_x]+O(p^2)\;\mbox{\rm with}\;
v_F={3\over2}tl.
\label{eq:linear}
\end{equation}
The Fermi velocity $v_F\approx10^6$ms$^{-1}$ in graphene. Now define modified
electron operators $a_\pm(\vec p)=a(\vec K_\pm+\vec p)$ etc, and combine them
into a four-spinor $\Psi=(b_+,a_+,a_-,b_-)^{tr}$. To linear order in $p$ the
complex structure implicit in (\ref{eq:linear}) results in the Dirac Hamiltonian
\begin{equation}
H=v_F\sum_{\vec p}\Psi^\dagger(\vec p)\vec\alpha.\vec p\,\Psi(\vec p), 
\end{equation}
with $4\times4$ matrices obeying $\{\alpha_i,\alpha_j\}=\delta_{ij}$. In other
words, low-energy excitations in graphene are massless relativistic fermions
with velocity $\approx c/300$. For monolayer graphene, the number of
relativistic flavors $N_f=2$, ie 8 spinor degrees of freedom = 2 C atoms/unit
cell $\times$ 2 Dirac points $\times$ 2 spins.

Electron-electron interactions in graphene can't be ignored for two reasons.
Firstly, Debye screening is suppressed due to the vanishing 
density of free electron states at the Dirac points. Secondly, the effective
fine structure constant is boosted by a factor $c/v_F$ and is hence of order
unity. The following effective field theory treats interactions in a simple
way~\cite{Khveshchenko1,Son:2007ja}:
\begin{equation}
S=\sum_{a=1}^{N_f}\int dx_0d^2x(\bar\psi_a\gamma_0\partial_0\psi_a
+v_F\bar\psi_a\vec\gamma.\vec\nabla\psi_a+iV\bar\psi_a\gamma_0\psi_a)\nonumber
+{1\over{2e^2}}\int dx_0d^3x(\partial_i V)^2.
\label{eq:action}
\end{equation}
The field $V$ is an ``instantaneous'' Coulomb potential governed by 3$d$ Maxwell
electrodynamics with $v_F\ll c$, interacting with relativistic electrons moving in the plane;
this is a ``braneworld''. In the large-$N_f$ limit the $V$-propagator is given
by
\begin{equation}
D(p)
=\left({{2\vert\vec p\vert}\over e^2}+{N_f\over8}{{\vert\vec
p\vert^2}\over{(p_0^2+v_F^2\vert\vec p\vert^2)^{1\over2}}}\right)^{-1}.
\end{equation} 
The first term is the classical result, the second is the leading quantum
correction arising from vacuum polarisation due to virtual electron-hole pairs.
In the static limit $p_0=0$ both contributions yield a $1/r$ potential. The
dimensionless combination
$\lambda=e^2N_f/16\varepsilon\varepsilon_0\hbar v_F\simeq1.4N_f/\varepsilon$
parametrises the relative strengths of quantum vs. classical, and depends on the
dielectric constant $\varepsilon$ of the substrate on which the graphene sits.
$\lambda$ is maximal for ``suspended'' graphene for which $\varepsilon=1$.

For sufficiently large $e^2/\varepsilon$ or sufficiently small $N_f$ it has been
hypothesised that the Fock vacuum is unstable with respect to condensation of
particle-hole pairs with $\langle\bar\psi\psi\rangle\not=0$, spontaneously
breaking the global U($2N_f$) symmetry of (\ref{eq:action}) to
U($N_f)\otimes$U($N_f)$ and leading to a gapping of the electron dispersion at
the Dirac points~\cite{Son:2007ja}. 
In particle physics we say a fermion mass is dynamically
generated via a chiral condensate; 
in condensed matter physics the resulting phase is known as a
{\it Mott insulator\/}. 
The transition occuring at $e_c^2(N_f)$ defines
a QCP whose universal properties characterise the low-energy excitations of
graphene. In the $(e^2,N_f)$ plane the QCPs lie along a phase boundary between
insulating and metallic phases, and the values of exponents such as $\delta$ in
the critical scaling relation 
\begin{equation}
\langle\bar\psi\psi\rangle\vert_{e^2=e_c^2}\propto m^{1\over\delta},
\end{equation}
where $m$ is an explicit symmetry-breaking mass parameter,
depend on $N_f$. This behaviour is similar to that of the 3$d$ Thirring model,
whose phase diagram has been mapped numerically over many years~\cite{Thirring}, and inspires a
treatment based on the technically simpler action (with units $v_F=1$)
\begin{equation}
S=\sum_{a=1}^{N_f}\int dx_0d^2x\left[\bar\psi_a\gamma_\mu\partial_\mu\psi_a
+iV\bar\psi_a\gamma_0\psi_a+{1\over{2g^2}}V^2\right]
\end{equation}
This has the identical $D(p)$ in the strong-coupling limit, but the Coulombic
$r^{-1}$ tail is
screened for $g^2<\infty$.
The relation between $g^2$ and $e^2,\lambda$ is not known {\it a priori\/}.

\section{Lattice Approach}
Due to the absence of small parameters at the QCP, a non-perturbative approach
is needed. We have used a lattice model based on the staggered fermion
formulation on a cubic lattice, with action ($i=1,\ldots,N$)~\cite{Hands:2008id}
\begin{equation}
S_{latt}={1\over2}\sum_{x\mu i}
\bar\chi^i_x\eta_{\mu x}(1+i\delta_{\mu0}V_{
x})\chi^i_{x+\hat\mu}
-\bar\chi^i_x\eta_{\mu x}(1-i\delta_{\mu0}V_{
x-\hat0})\chi^i_{x-\hat\mu}
+m\sum_{xi}\bar\chi^i_x\chi^i_x+{N\over4g^2}\sum_{x}V_x^2
\label{eq:Slatt}
\end{equation}
The fermion fields $\chi$, $\bar\chi$ are defined on lattice sites, and the
boson fields $V$ on the timelike links. The sign factors $\eta_{\mu
x}\equiv(-1)^{x_0+\cdots+x_{\mu-1}}$ ensure a covariant continuum limit in the
free-field limit $g^2=0$. The action (\ref{eq:Slatt}) only has a chance of
recovering the physics of (\ref{eq:action}) at a QCP, where for instance details
of the underlying lattice should become irrelevant. However, (\ref{eq:Slatt})
exhibits a distinct chiral symmetry breaking pattern
U($N)\otimes$U($N)\to$U$(N)$; away from weak coupling, there is no guarantee of
``taste symmetry restoration'' ensuring the correct continuum symmetries with
$N_f=2N$.

\begin{figure}[t]
  \hfill
  \begin{minipage}[t]{.49\textwidth}
    \begin{center}
\includegraphics[width=0.89\textwidth]{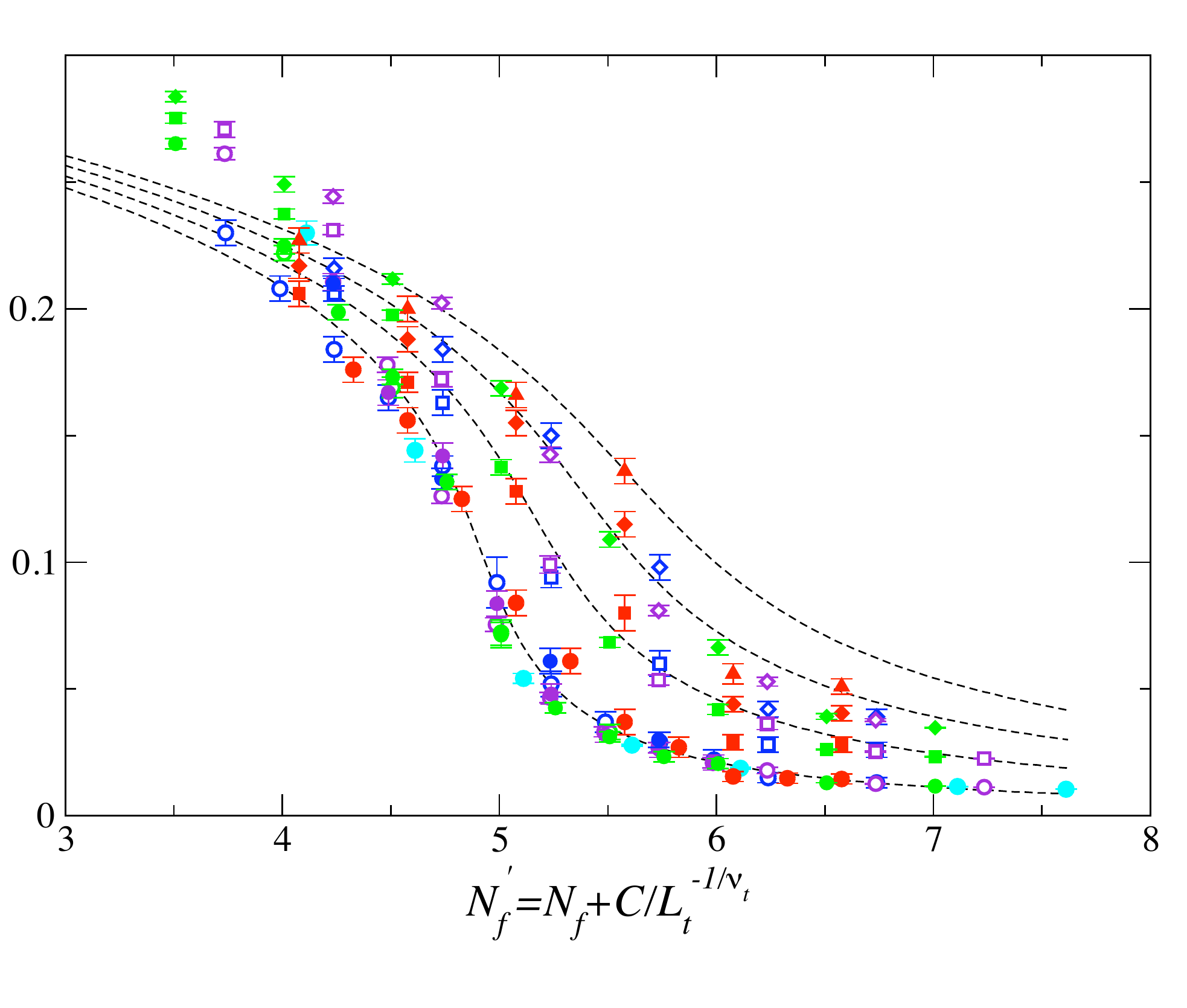}
    \end{center}
  \end{minipage}
  \hfill
  \begin{minipage}[t]{.49\textwidth}
    \begin{center}
\includegraphics[width=0.99\textwidth]{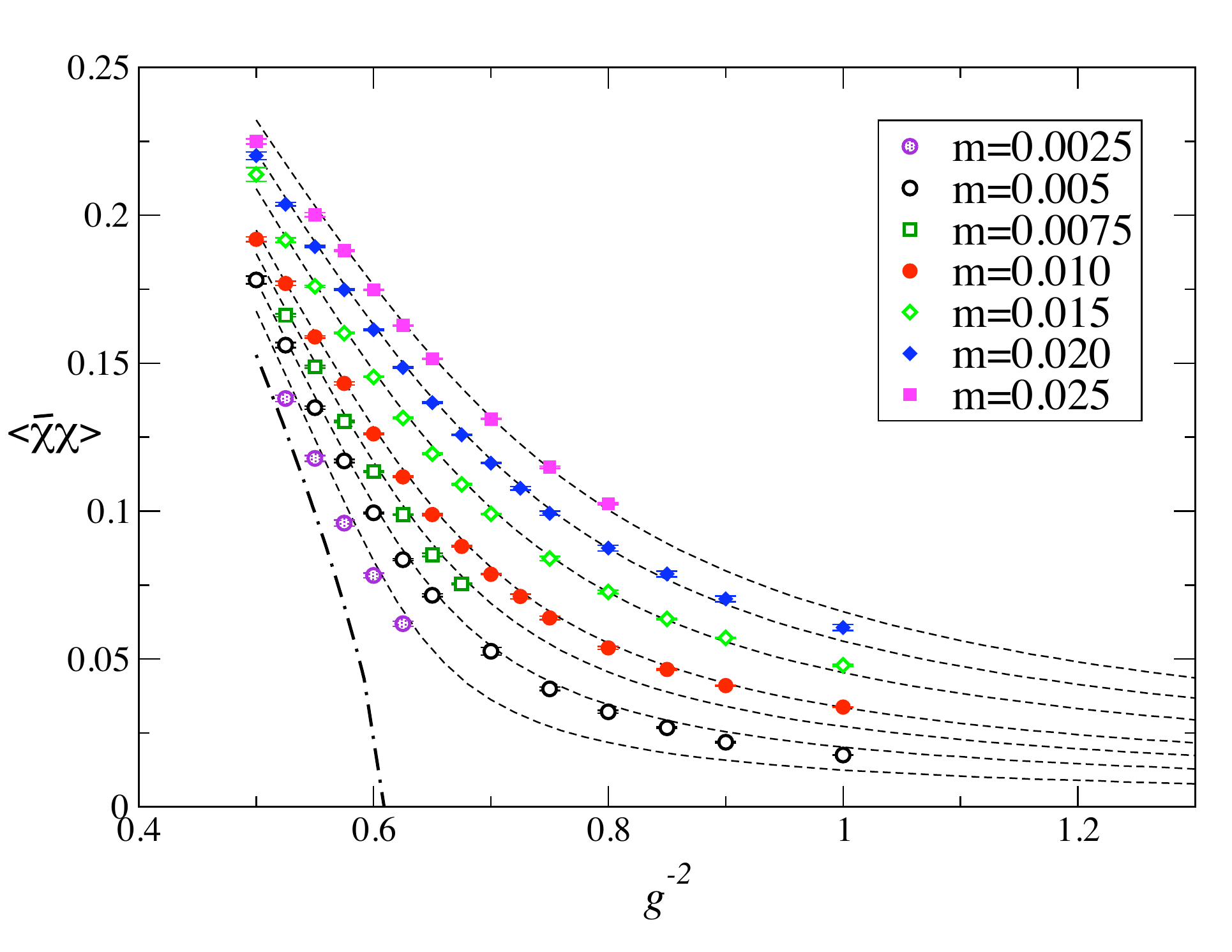}
    \end{center}
  \end{minipage}
  \hfill
\caption{Chiral order parameter $\langle\bar\chi\chi\rangle$ resulting from (2.1)
for various $m$ at
strong coupling (left), and for $N_f=2$ (right).} 
\label{fig:eosplots}
\end{figure}
Fig.~\ref{fig:eosplots} shows the order
parameter $\langle\bar\chi\chi(m)\rangle$ as a function of $N_f$ in the strong
coupling limit (left)~\cite{Hands:2008id}, and as a function of $g^{-2}$ for the monolayer value
$N_f=2$ (right)~\cite{Armour:2009vj}. In both cases a power-law equation of state appropriate for a
continuous phase transition has been fitted; this identifies  the location of the
QCP. The left hand plot shows a critical value $N_{fc}=4.8(2)$, proving that a
QCP exists for finite $g^2$ for $N_f=2<N_{fc}$, and is thus potentially relevant
for monolayer graphene. The right hand plot shows a critical coupling
$g_c^{-2}=0.609(2)$ for $N_f=2$, with the critical exponent $\delta=2.66(3)$, to
be contrasted with $\delta(N_{fc})=5.5(3)$. The model (\ref{eq:Slatt}) has also
been used to study the symmetry restoring transition at temperature $T>0$,
expected to be of BKT type~\cite{Armour:2011hf}.

These results are consistent with
those obtained by simulating the braneworld
gauge theory~\cite{D&L}. That work initially suggested the QCP might occur at a coupling
sufficiently small that suspended graphene samples might lie in the insulating
phase. Later simulations on a 2$d$ honeycomb lattice with a more realistic
inter-electron potential suggest the critical coupling is slightly too strong for a
Mott insulator to be observed, although modification of the electron dispersion
in pure suspended samples
due to the vicinity of a QCP is not excluded~\cite{Ulybyshev:2013swa}. 

\section{Bilayer Graphene}
There is just as much interest in graphene samples formed from two honeycomb
layers, in which the $A$ atoms of one sheet lie directly over the $B$ atoms of
its neighbour. Naively, electron transport should have an effective description with $N_f=4$
relativistic fermions; however in the absence of any further interactions the
Hamiltonian (\ref{eq:tight-binding}) actually predicts a parabolic band
structure~\cite{McCann}. 
Let's rename the original intralayer coupling along a CC bond within a layer $t_0$
and introduce new interlayer couplings $t_1$ between overlying $AB$ sites and
$t_3$ between $AB$ sites displaced horizontally a distance $l$. This latter
coupling results in a trigonal distortion of the parabolic band, breaking it
into four separate cones at the Dirac point. It turns out an $N_f=4$ effective
field theory description is plausible for long-wavelength excitations with 
$kl\lapprox t_1t_3/t_0^2$. 

The new ingredient in the bilayer problem is the possibility of applying a bias
voltage across the layers, inducing a negative charge on one layer due to
non-zero density of electrons, and a positive charge on the other due to holes.
Denoting this voltage by $2\mu$ we can introduce it with positive minimal
coupling to fields $\psi$ on the the upper layer and negative coupling to
fields $\phi$ on the lower. This is formally identical to the introduction of an
isopsin chemical potential in two-flavor QCD. The lagrangian density in
continuum notation reads~\cite{Armour:2013yk}
\begin{equation}
{\cal L}=(\bar\psi,\bar\phi)\left(\matrix{D[V;\mu]+m&ij\cr
-ij&D[V;-\mu]-m\cr}\right)\left(\matrix{\psi\cr\phi}\right)+{1\over{2g^2}}V^2
\equiv\bar\Psi{\cal M}\Psi+{1\over{2g^2}}V^2,
\label{eq:Sbilayer}
\end{equation}
where $D[V;\mu]=D[V;0]+\mu\gamma_0=-D^\dagger[V;-\mu]$. The parameter $m$ is a symmetry-breaking gap
parameter due to intralayer particle-hole pairing, and $j$ a gap parameter due
to interlayer pairing; the bound state formed in the latter case is known as an {\it
exciton\/}. They act as IR regulators for the model; in practice
numerical simulations with $m=0$, $j\not=0$ are perfectly feasible. The model is
not realistic in the sense that intralayer $\psi$ -- $\psi$ interactions have the same
coupling strength as interlayer $\psi$ -- $\phi$; however, from
a simulator's point of view the key feature is the identity
\begin{equation}
\mbox{\rm det}{\cal M}=\mbox{\rm det}[(D+m)^\dagger(D+m)+j^2]>0.
\end{equation}
Even with $\mu\not=0$ the model has no sign problem and is amenable to orthodox
Monte Carlo methods.

We have implemented a lattice version of (\ref{eq:Sbilayer}) using $N=2$
staggered fermion flavors on $32^3$ and $48^3$ lattices. Since $N_f=4$ is close
to $N_{fc}$, it is hard to identify the location of the QCP; we set
$(g^2a)^{-1}=0.4$\footnote{Care is needed since {\it a priori} there
is no reason for spatial and temporal lattice spacings to coincide at the QCP.}, 
which we believe is just to the sub-critical side of the QCP.
Simulations with varying $\mu$ have been performed, and the following
observables monitored:
\begin{eqnarray}
\mbox{carrier density}\;n_c&\equiv&{{\partial\ln{\cal Z}}\over{\partial\mu}}=
\langle\bar\psi D_0\psi\rangle-\langle\bar\phi D_0\phi\rangle,\\
\mbox{exciton condensate}\;\langle\Psi\Psi\rangle&\equiv&{{\partial\ln{\cal
Z}}\over{\partial j}}=
i\langle\bar\psi\phi-\bar\phi\psi\rangle,\label{eq:excon}\\
\mbox{chiral condensate}\;\langle\bar\Psi\Psi\rangle&\equiv&{{\partial\ln{\cal
Z}}\over{\partial m}}=
\langle\bar\psi\psi-\bar\phi\phi\rangle.
\end{eqnarray}
The exciton condensate $\langle\Psi\Psi\rangle$ (\ref{eq:excon}) caused by spontaneous pairing of
electrons and holes between layers is a new feature; just like any other pairing
phenomenon it is expected to induce a gap rendering the ground state insulating.
By varying $\mu$ we can check how exciton condensation varies with the extent of
the Fermi surface (strictly a Fermi line in 2+1$d$).

\begin{figure}[t]
  \hfill
  \begin{minipage}[t]{.49\textwidth}
    \begin{center}
\includegraphics[width=0.99\textwidth]{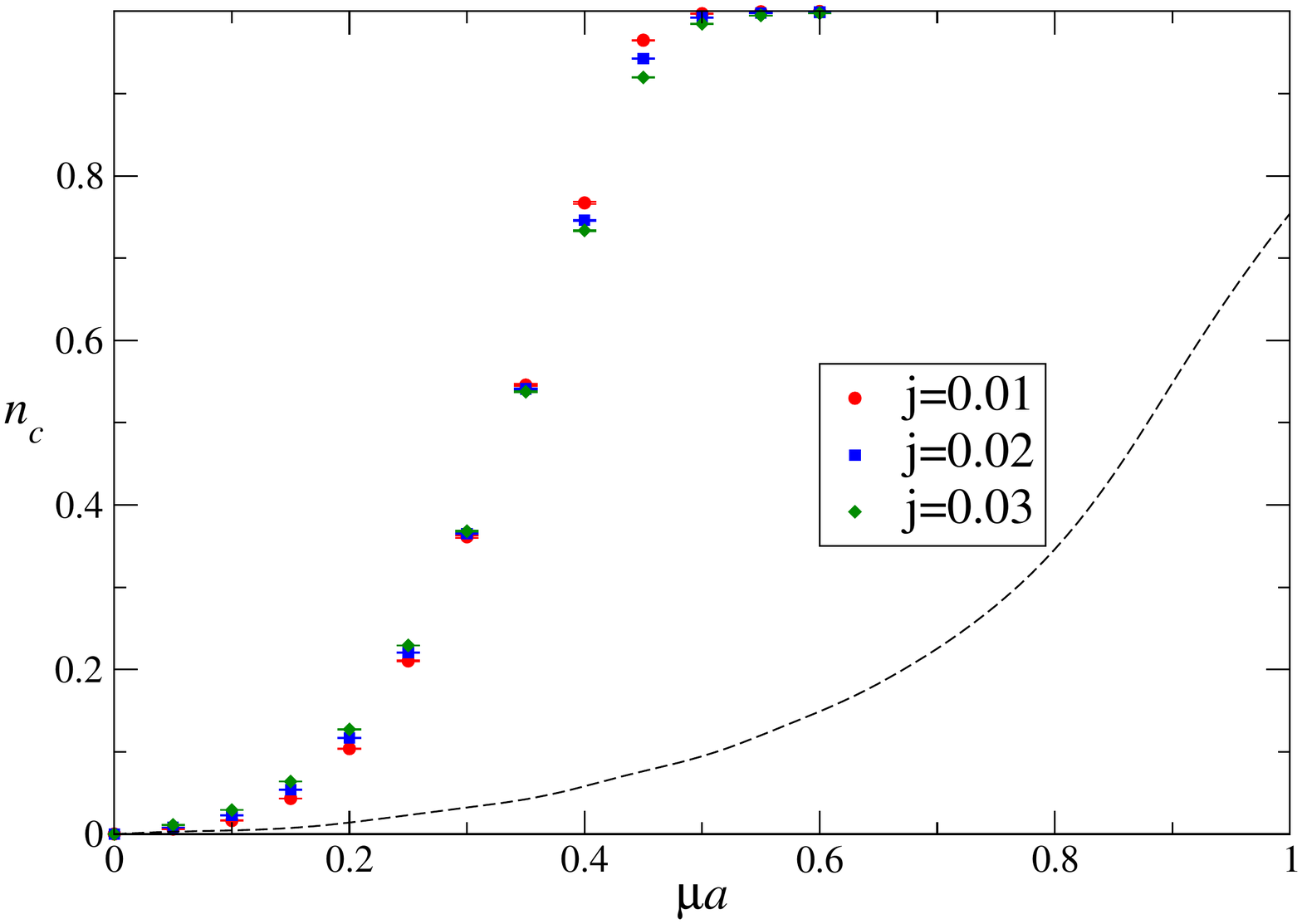}
    \end{center}
  \end{minipage}
  \hfill
  \begin{minipage}[t]{.49\textwidth}
    \begin{center}
\includegraphics[width=0.905\textwidth]{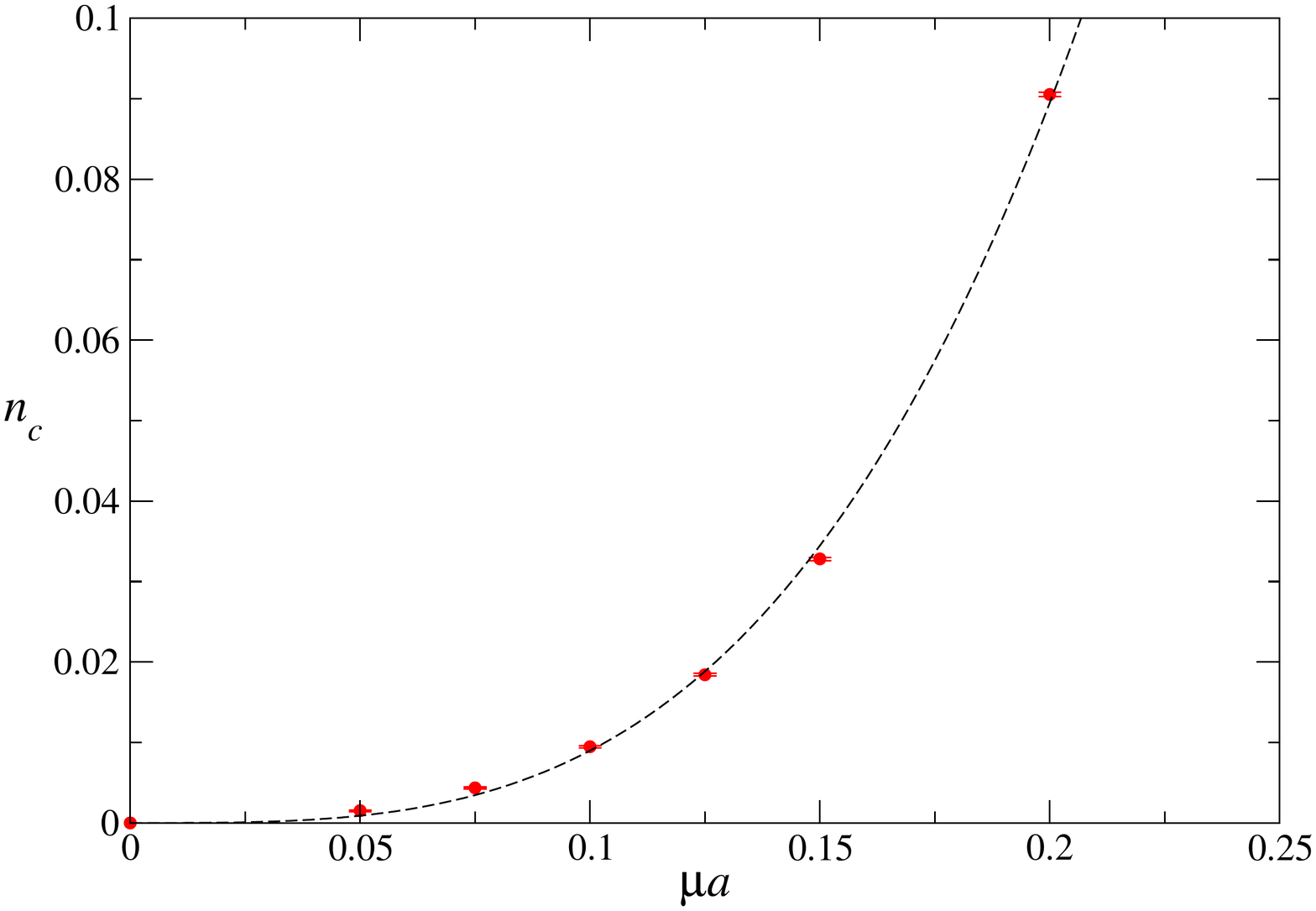}
    \end{center}
  \end{minipage}
  \hfill
\caption{Carrier density $n_c$ in bilayer model as a function of $\mu$ 
(left), and in the small $\mu$ $j\to0$ limit (right).} 
\label{fig:carrierdensity}
\end{figure}
Fig.~\ref{fig:carrierdensity} shows simulation results for $n_c$; at the left the data rise
steeply with $\mu$ to reach a plateau at $\mu a\approx0.5$. There is no
discernable onset $\mu_o$ below which $n_c$ vanishes. The plateau marks the {\it
saturation region\/} where the lattice contains one fermion (both particle and
hole) per site and the Exclusion Principle prevents further occupation. It should
be regarded as an artifact of simulating continuous fields on a discrete
lattice. Interestingly, all other known simulable lattice models reach
saturation at $\mu a\sim O(1)$; indeed, the result for free fermions in 2+1$d$
is shown as a dashed line. In the limit $T\to0$, $\mu$ coincides with the Fermi
energy $E_F$. Since $n_c$ is directly related to the Fermi momentum $k_F$, the
precocious saturation is the first hint that $E_F<k_F$, ie. the system is
strongly self-bound.
At right the $j\to0$ extrapolated data are fitted to a power law, with the
result $n_c(j=0)\propto\mu^{3.32(1)}$. The density rises faster than the
free-field expectation $n_c\propto\mu^2$ coming from the area of the Fermi disk.

\begin{figure}[t]
  \hfill
  \begin{minipage}[t]{.49\textwidth}
    \begin{center}
\includegraphics[width=0.99\textwidth]{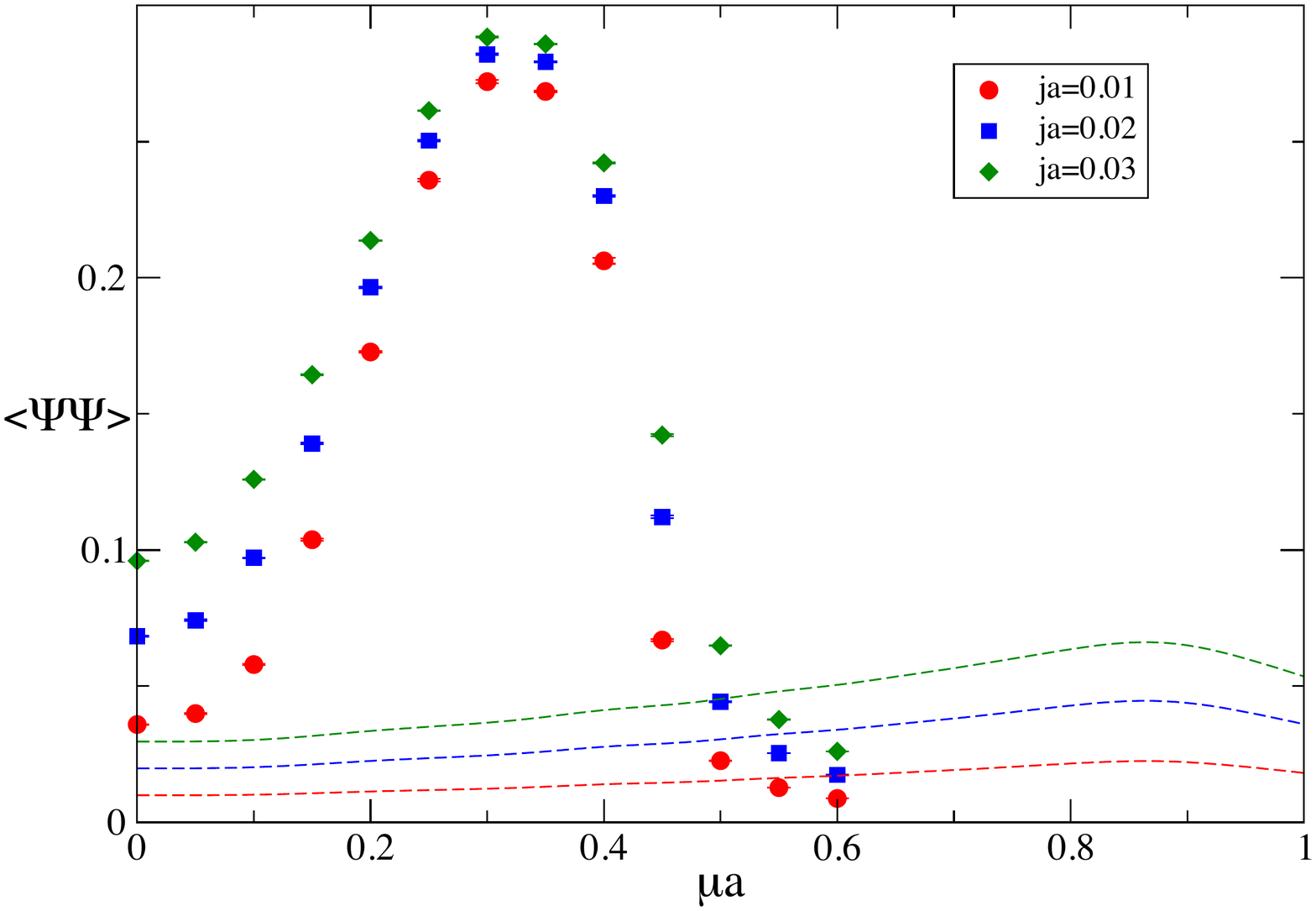}
    \end{center}
  \end{minipage}
  \hfill
  \begin{minipage}[t]{.49\textwidth}
    \begin{center}
\includegraphics[width=0.905\textwidth]{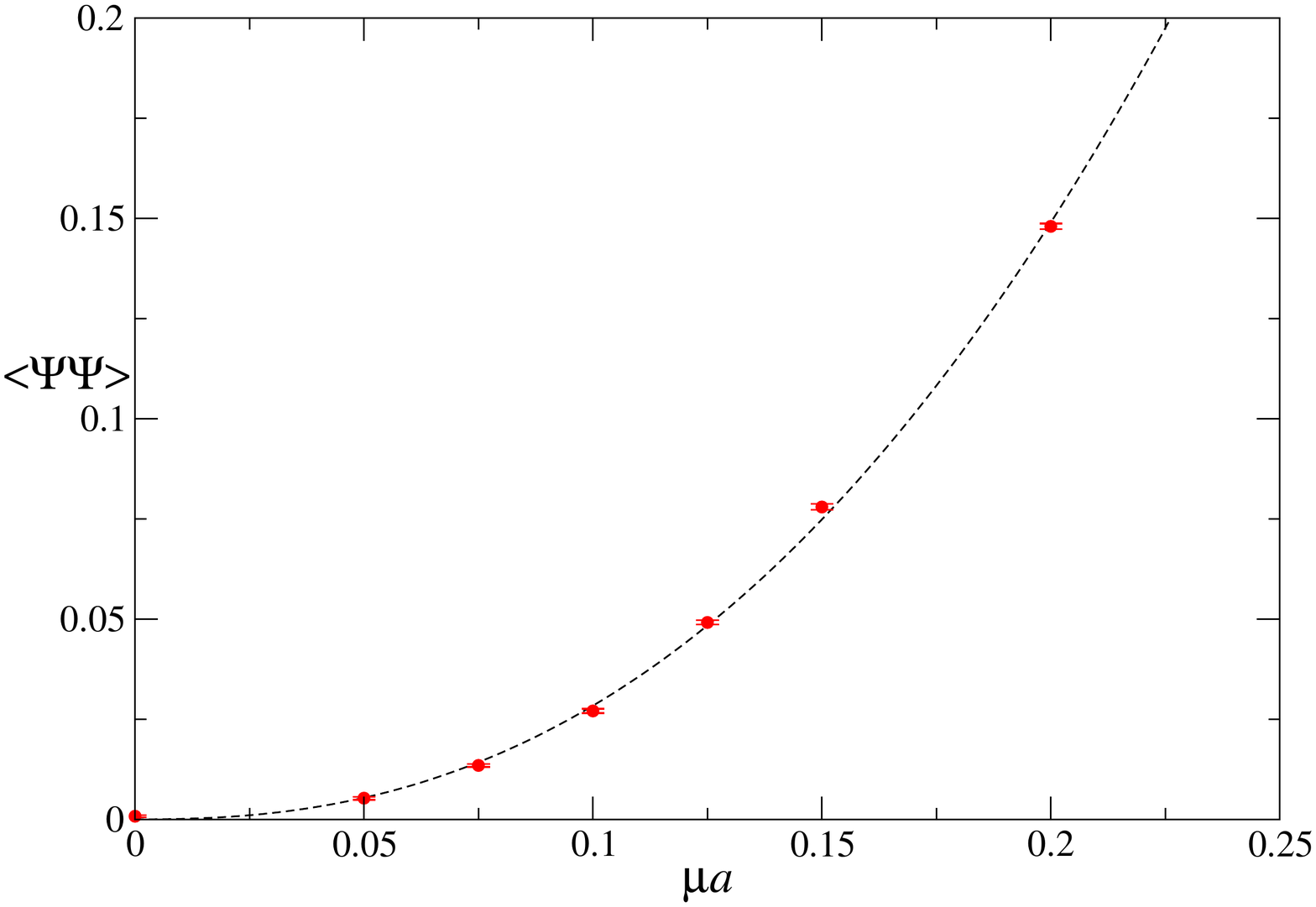}
    \end{center}
  \end{minipage}
  \hfill
\caption{Exciton condensate $\langle\Psi\Psi\rangle$ as a function of $\mu$ 
(left), and in the small $\mu$ $j\to0$ limit (right).} 
\label{fig:exciton}
\end{figure}
Fig.~\ref{fig:exciton} shows the corresponding plots for
$\langle\Psi\Psi\rangle$. Again there is a steep rise to a peak at $\mu
a\approx0.3$, followed by an even more rapid descent to zero in the saturation
region. The free-field results, which are much smaller and extremely sensitive to $j$, are shown
as dashed lines. The extrapolated data is fitted by
$\langle\Psi\Psi(j=0)\rangle\propto\mu^{2.39(2)}$. 
As we will see below, this result also
suggests a weak-coupling approach is not applicable.

\begin{figure}[thb]
  \hfill
  \begin{minipage}[t]{.49\textwidth}
    \begin{center}
\includegraphics[width=0.91\textwidth]{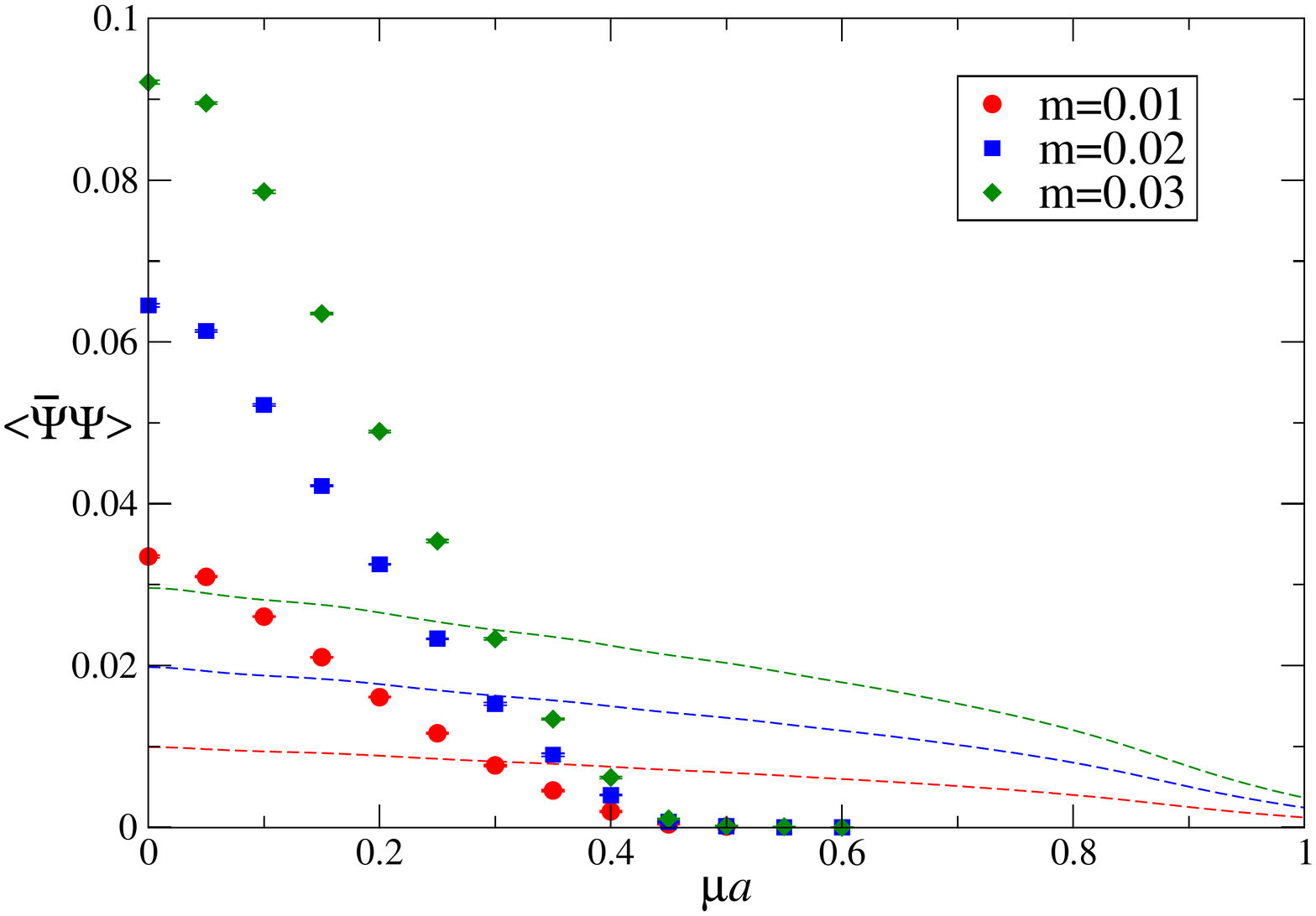}
    \end{center}
  \end{minipage}
  \hfill
  \begin{minipage}[t]{.49\textwidth}
    \begin{center}
\includegraphics[width=0.99\textwidth]{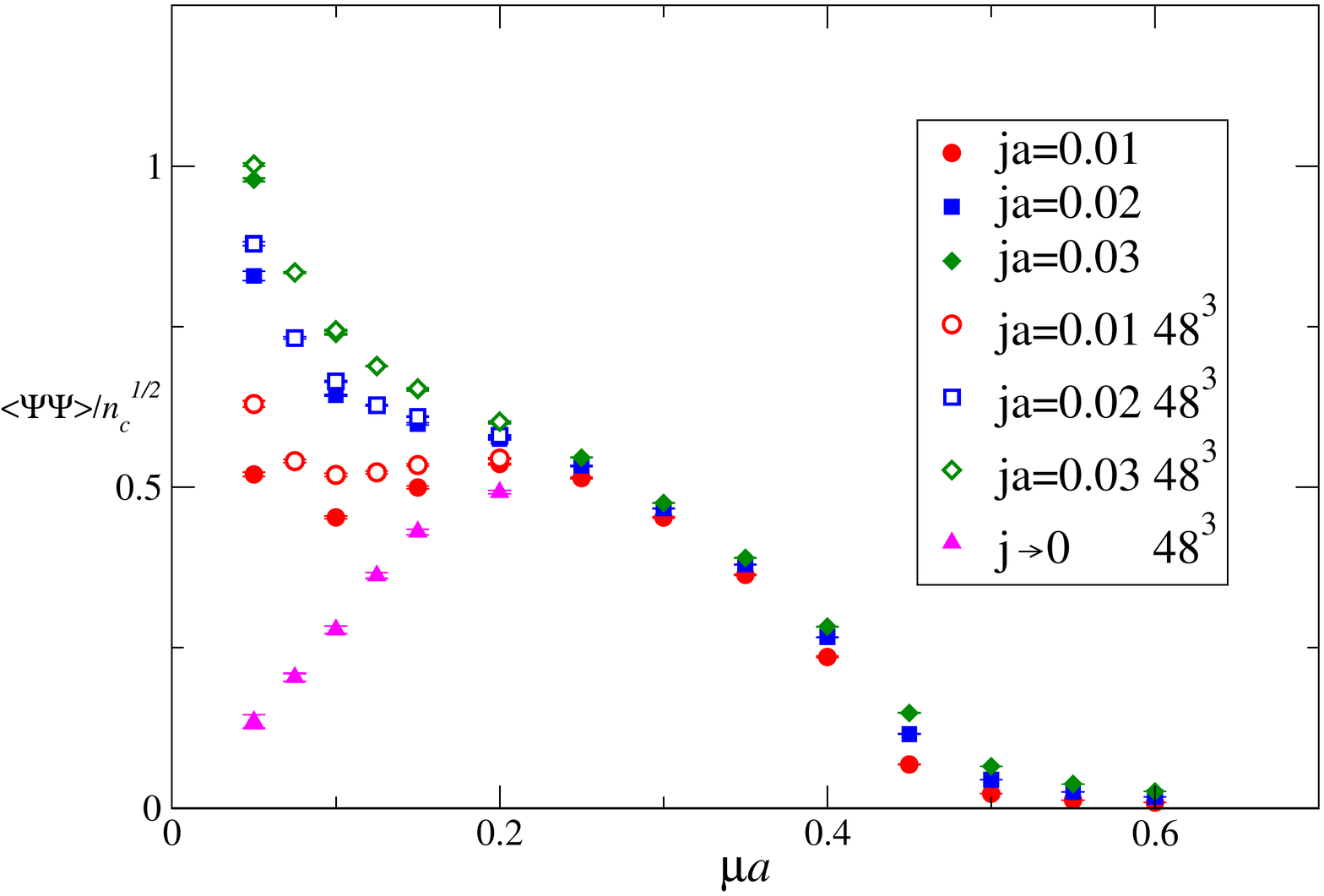}
    \end{center}
  \end{minipage}
  \hfill
\caption{Chiral condensate $\langle\bar\Psi\Psi\rangle$ vs. $\mu$ 
(left), and the ratio $\langle\Psi\Psi\rangle/\surd n_c$ vs $\mu$ for various
$j$ (right).} 
\label{fig:conrat}
\end{figure}
The left panel of Fig.~\ref{fig:conrat} shows the intralayer condensate
$\langle\bar\Psi\Psi\rangle$; for small $\mu$ it exceeds its free-field value,
as might be expected from the strong interactions close to a QCP, but then
rapidly falls to zero as $\mu$ increases, so that by the time $\mu a$=0.3 where
$\langle\Psi\Psi\rangle$ peaks, it falls below the free-field value. This
can be understood in terms of a competition between the two condensates; intralayer pairing
is suppressed as $E_F$ grows, because more energy is needed to excite a
particle-hole pair from a single layer, whereas interlayer pairing is enhanced
by the larger density of states as $k_F$ grows. Even at $\mu=0$
$\vert\langle\bar\Psi\Psi\rangle\vert\approx{1\over3}\vert\langle\Psi\Psi\rangle\vert_{peak}$.

Now, $\langle\Psi\Psi\rangle$ is
simply the density of exciton pairs in the ground state. In a weakly-coupled BCS
picture
of exciton condensation, the excitons would be drawn from a shell of thickness
$\Delta$ around the Fermi surface, where $\Delta$ is the induced gap; we would
then expect $\langle\Psi\Psi\rangle\propto\Delta k_F\propto\Delta
n_c^{1\over2}$, where the last step follows from Luttinger's theorem relating
$n_c$ to the volume enclosed within the Fermi surface. Thus we predict 
\begin{equation}
\Delta(\mu)\propto{{\langle\Psi\Psi(\mu)\rangle}\over{\surd n_c(\mu)}}.
\label{eq:ratio}
\end{equation}
The quantity on the RHS of (\ref{eq:ratio}) is plotted as a function of $\mu$ in the RH panel of
Fig.~\ref{fig:conrat}, with particular attention paid to the extrapolation
to both thermodynamic and $j\to0$ limits. For small $\mu$ we infer
$\Delta\propto\mu$; at a QCP no other behaviour could arise, since $\mu$ is the
only scale. This should be contrasted with other models studied using lattice
simulation; in the non-renormalisable NJL model, $\Delta\propto\Lambda_{\rm
UV}$~\cite{Hands:2004uv},
whereas in two-color QCD $\Delta=O(\Lambda_{\rm QCD})$~\cite{Cotter:2012mb}. In neither case is there
any significant dependence on $\mu$.

\section{Quasiparticle Dispersion in Bilayer Graphene}
In order to confirm the picture developed in the previous section of a system
with $k_F>\mu$ and $\Delta\propto\mu$ it would clearly help to have direct
knowledge of these quantities from the fermion (in this context known as the {\it
quasiparticle\/}) dispersion relation $E(k)$. Here we present preliminary
results obtained from the Euclidean propagator
$\langle\Psi(k)\bar\Psi(k)\rangle\propto e^{-E(k)t}$ on $32^3$; to improve
momentum resolution partially twisted boundary conditions are
used~\cite{Flynn:2005in}, so that the
smallest non-zero $k$ is $\pi/48a$.
\begin{figure}[thb]
  \hfill
  \begin{minipage}[t]{.49\textwidth}
    \begin{center}
\includegraphics[width=0.99\textwidth]{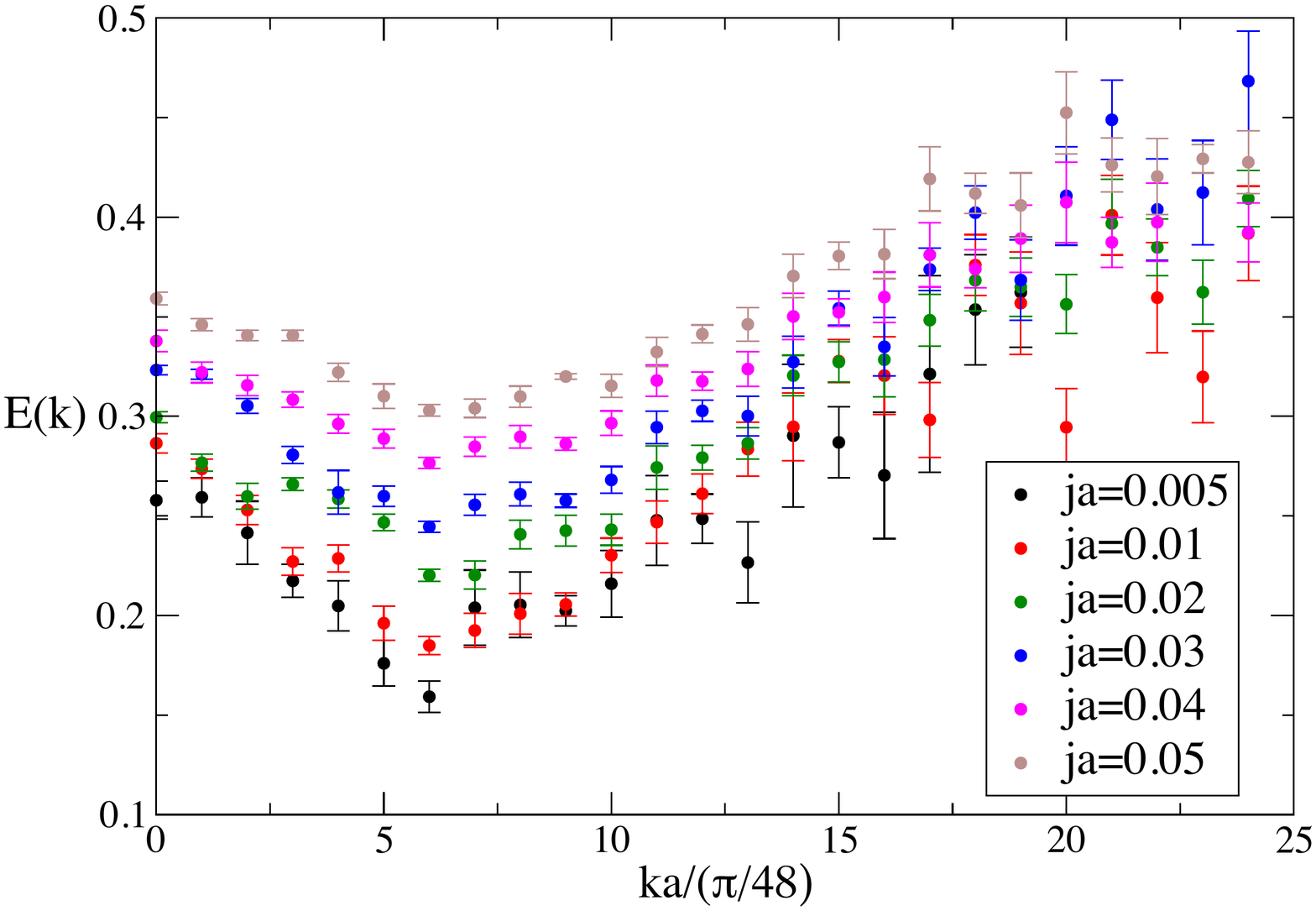}
    \end{center}
  \end{minipage}
  \hfill
  \begin{minipage}[t]{.49\textwidth}
    \begin{center}
\includegraphics[width=0.99\textwidth]{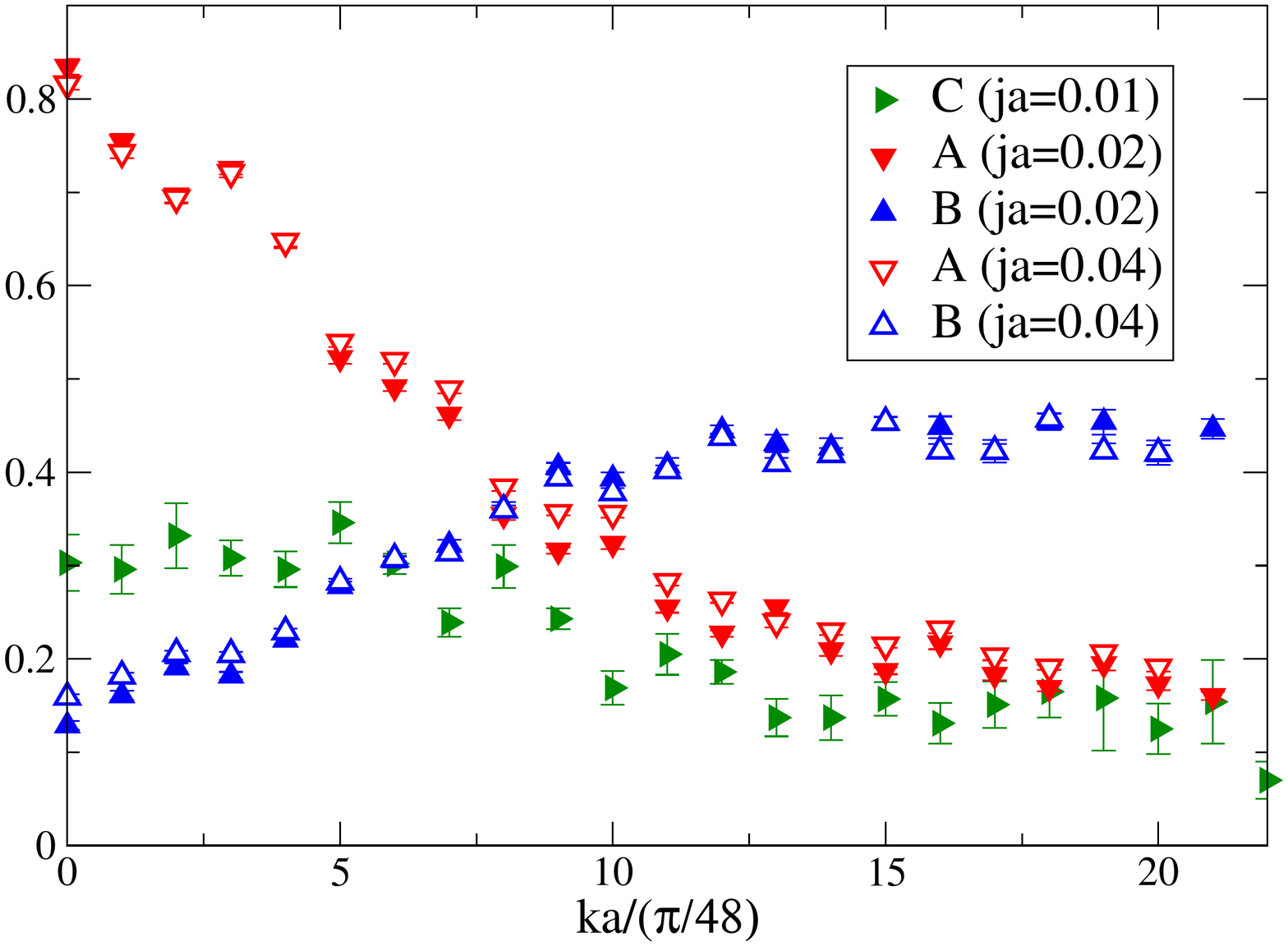}
    \end{center}
  \end{minipage}
  \hfill
\caption{Quasiparticle dispersion relation $E(k)$ at $\mu a=0.2$ for various
$j$ (left), 
and the fitted propagator amplitudes (right).}
\label{fig:dispersion}
\end{figure}
We calculated the timeslice correlators in both normal and anomalous channels
according to
\begin{equation}
C_N(\vec k,t)=\sum_{\vec x}\langle\psi(\vec0,0)\bar\psi(\vec x,t)\rangle
e^{-i\vec k.\vec x};\;\;
C_A(\vec k,t)=\sum_{\vec x}\langle\psi(\vec0,0)\bar\phi(\vec x,t)\rangle
e^{-i\vec k.\vec x}
\end{equation}
and fitted to the forms~\cite{Hands:2001aq}.
\begin{equation}
C_N(k,t)=Ae^{-Et}+Be^{-E(L_t-t)};\;\;C_A(k,t)=C(e^{-Et}-e^{-E(L_t-t)}).
\end{equation}
The resulting dispersion $E(k)$ is shown for several $j$ in the left panel of
Fig.~\ref{fig:dispersion}. The clear dip marks the Fermi surface; for
$k<k_F$ the correlator is dominated by forwards-propagating holes, and for
$k>k_F$ by backwards-propagating particles, as demonstrated in the relative
magnitudes of the amplitudes $A$ and $B$ in Fig.~\ref{fig:dispersion} (right).
Note that the amplitude $C$ for anomalous propagation, in which eg. an electron in
one layer is absorbed by an exciton transferring its momentum to an electron in
the other, is comparable in magnitude to those for normal propagation for a
range of momenta with $k\approx k_F$. 

From the plots we estimate $k_Fa\approx0.4>\mu a=0.2$, supporting the earlier
contention that $k_F>E_F$ due to the self-bound nature of the system
near the QCP. As a sanity check we note that for free lattice
fermions the carrier density $n_c^{freelatt}(\mu a=0.4)a^2\approx0.06$, to be
compared with the directly measured value 0.09 (see
Fig.~\ref{fig:carrierdensity}).

\begin{figure}[thb]
\begin{centering}
\includegraphics[width=.5\textwidth]{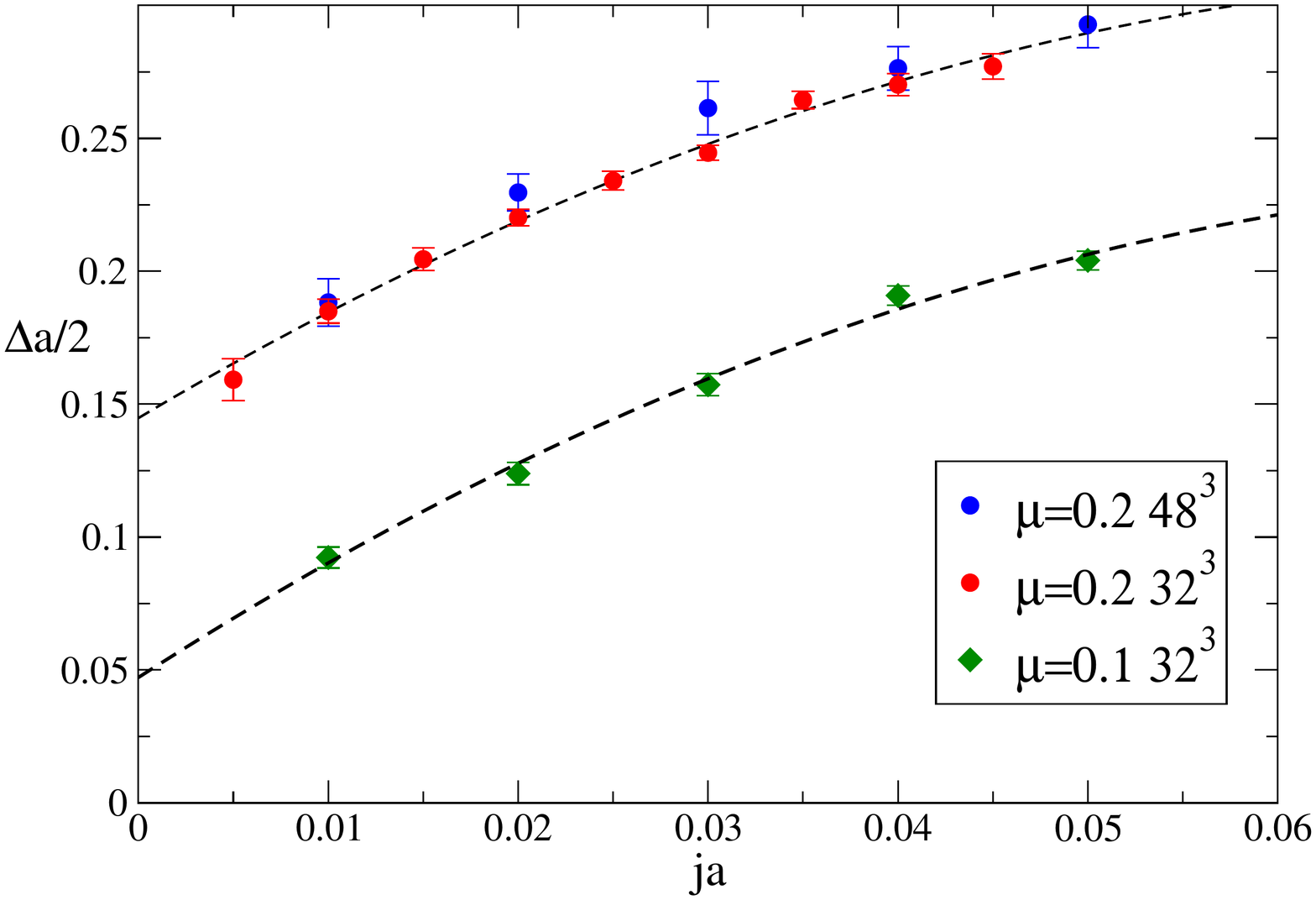}
\caption{Gap $\Delta(j)$ estimated from the minimum of $E(k)$.}
\label{fig:gap}
\end{centering}
\end{figure}
Finally in Fig.~\ref{fig:gap} we plot the gap $\Delta$ estimated from the minima
of $E(k)$ in Fig.~\ref{fig:dispersion}, with data taken at two different values
of $\mu$. We can then extrapolate to $j=0$ and estimate $\Delta a(\mu
a=0.1)\approx0.1$, $\Delta a(\mu=0.2)\approx0.3$. This is roughly consistent
with our earlier prediction that $\Delta\propto\mu$, and supports a physical
picture of a gapped Fermi surface with $\Delta/\mu\sim O(1)$. This should be
compared with studies using a diagrammatic approach predicting
$\Delta/\mu\sim10^{-7}$~\cite{Kharitonov}.

\section{Summary}
We've discussed how the effective description of graphene in terms of
relativistic field theory can be studied non-perturbatively using orthodox
lattice simulation techniques familiar from particle physics. The simulations
have confirmed at least the theoretical possibility of a QCP in monolayer
graphene which may have relevance for the correct description of charge transport in 
suspended samples. In the case of bilayer graphene the model presented here,
though arguably not very realistic due to the equality of intra- and inter-layer
interaction stengths, is at least an interesting new member of the rather exclusive stable of
models which permit study by Monte Carlo techniques in the presence of a chemical
potential. As we have seen, its behaviour is rather different from other models
in this class (NJL, QC$_2$D) because residual interactions at the Fermi surface
are strong in the vicinity of a QCP.

\end{document}